\documentclass[reprint, amsmath,amssymb,aps]{revtex4-1}

\usepackage{graphicx}      
\graphicspath{{./ProfiPics/},{./Graphs/v13/},{./Pics/}}
\usepackage{bmpsize}
\usepackage{braket} 
\usepackage{bbold}
\usepackage[ansinew]{inputenc} 
\usepackage{hyperref}

\usepackage{color}

\begin{document}


\title{Real-time monitoring of L\'evy flights in a single quantum system}
\author{M. Issler}
\author{J. H\"oller}
\author{A. Imamo\u{g}lu}
\affiliation{Institute of Quantum Electronics, ETH Zurich, CH-8093
Zurich, Switzerland.}


\author{}
\affiliation{}


\date{\today}

\begin{abstract}
L\'evy flights are random walks where the dynamics is dominated by rare events. Even though they have been studied in vastly different physical systems, their observation in a single quantum system has remained elusive. Here we analyze a periodically driven open central spin system and demonstrate theoretically that the dynamics of the spin environment exhibits L\'evy flights. For the particular realization in a single-electron charged quantum dot driven by periodic resonant laser pulses, we use Monte Carlo simulations to confirm that the long waiting times between successive nuclear spin flip events are governed
by a power-law distribution; the corresponding exponent $\eta =-3/2$ can be directly measured in real-time by observing the
waiting time distribution of successive photon emission events. Remarkably, the dominant intrinsic limitation of the scheme arising from nuclear quadrupole coupling can be minimized by adjusting the magnetic field or by implementing spin echo.
\end{abstract}


\maketitle

L\'evy flights are continuous-time random walks that are characterized by diverging average waiting times or average step lengths. L\'evy flights occur if either the waiting time $\Delta t$ or the step length $\Delta x$ distribution show a power-law dependence, $P(\Delta t)\sim \Delta t^{\eta}$ or $P(\Delta x)\sim \Delta x^{\eta}$, with an exponent $-2 < \eta < -1$. The fat tail of the distribution leads to extreme events that dominate the dynamics \cite{shlesinger93}. L\'evy flights have been observed experimentally in various physical systems, such as light scattering in specially engineered materials \cite{barthelemy08} or in subrecoil laser cooling \cite{bardou94}. In all experiments to date, L\'evy dynamics had to be inferred from ensemble properties; tracking L\'evy flights of a single quantum system in real time has so far remained elusive.

In this Letter, we theoretically investigate an open central spin system driven by periodic laser pulses and show that the random walk of the polarization of the spin environment can be characterized by L\'evy flights. To be specific, we derive the dynamics for a particular realization of the central spin model: an electron confined in a single self-assembled quantum dot (QD) that interacts with the nuclear spin environment. This choice is motivated by striking experimental results on an ensemble of self-assembled QDs which are compatible with L\'evy flights of the Overhauser field \cite{greilich06, bayer07,barnes11, economou14, yugova09}. The random walker in this central spin system corresponds to the nuclear Overhauser field $B_{\mathrm{Oh}}$,
the step length to the shift in the Overhauser field $\Delta B_{\mathrm{Oh}}$ induced by a nuclear spin flip and the waiting time is given by the time interval $\Delta t_{\mathrm{nuc}}$ between two successive nuclear spin flips. Nuclear spin flips in this system are enabled by spontaneous emission from the laser excited central (electron) spin. Conversely, long waiting times are characterized by a corresponding reduction in the photon emission rate. We confirm that the waiting time has a power-law distribution with an exponent $\eta=-3/2$ leading to extremely long waiting times. These L\'evy flights can be observed by monitoring the waiting time distribution of successive photon detection events. Owing to the favorable ratio of the rates of emitted photons to nuclear spin flips it is experimentally feasible to probe the nuclear spin dynamics in real-time even with a modest photon detection efficiency.

\begin{figure}
 \includegraphics[natwidth=8.6cm, natheight=9.5cm]{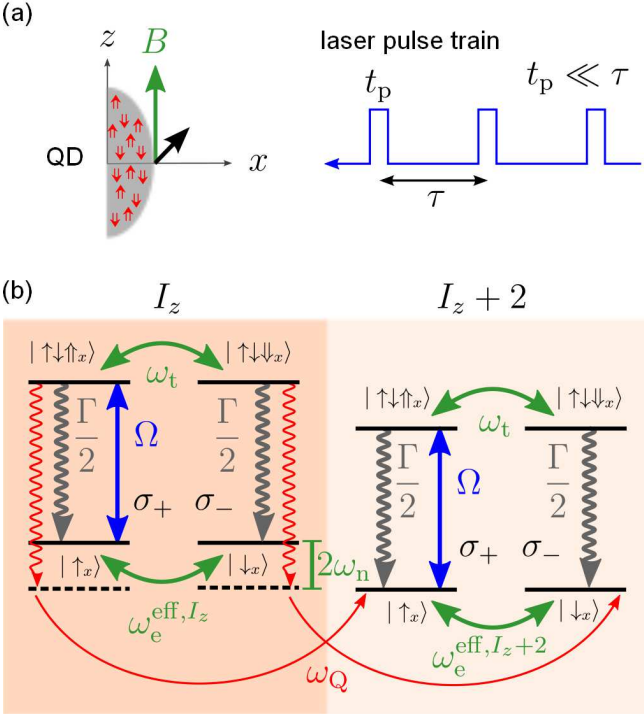}
 \caption{(a) The magnetic field $B$ is applied in $z$-direction and the laser pulse train propagates along the $x$-direction, which is the QD growth axis.
 (b) The electron and the trion precess in the external magnetic field (green) and the $\sigma_+$-polarized laser only couples to $\ket{\uparrow_x}$ (blue). Spontaneous emission is measured in the circularly polarized basis (gray). On one hand, the coupling to the nuclear spin environment modifies the electron ground state precession frequency, while on the other hand the slow nuclear spin dynamics is dominated by optically-assisted nuclear spin flips (red).
\label{fig:coupling}}
\end{figure}

We consider an electron in a single self-assembled QD coupled to the radiation field reservoir and illuminated by a ($\sigma^+$) circularly polarized laser pulse train propagating along the QD growth
direction $x$. The external magnetic field $\vec{B} = B \hat{\vec{z}}$ is applied transverse to the light propagation direction (Voigt geometry) as shown in FIG. \ref{fig:coupling} (a). The electronic ground states
$(\ket{ \uparrow_x }, \ket{\downarrow_x}$) and the excited, negatively-charged trion states ($\ket{\uparrow \downarrow\Uparrow_x}, \ket{\uparrow \downarrow\Downarrow_x}$)
realize a four-level system with ground ($\mathrm{e}$) and excited ($\mathrm{t}$) states precessing in the external magnetic field $B$ with frequencies $\omega_{\mathrm{e}} = g_{\mathrm{e}} \mu_{\mathrm{B}} B$ and
$\omega_{\mathrm{t}} = g_{\mathrm{h}} \mu_{\mathrm{B}} B$, respectively. The effective g-factors of the electron (hole) are given by $g_{\mathrm{e}} < 0$ ($g_{\mathrm{h}} > 0$), and $\mu_{\mathrm{B}}$ is the Bohr magneton.
We describe the system in the basis of the light propagation direction and hereafter suppress the subscript $x$ on the basis states, i.e. $\ket{\uparrow_x} = \ket{\uparrow}$.
The Hamiltonian is then given by ($\hbar = 1$)
\begin{eqnarray*}
\hat{H}_0 =  && +\frac{\omega_{\mathrm{e}}}{2} (\ket{ \uparrow } \bra{ \downarrow } + \ket{ \downarrow } \bra{ \uparrow }) \\
&&+ \frac{\omega_{\mathrm{t}}}{2} ( \ket{ \uparrow \downarrow\Uparrow} \bra{ \uparrow \downarrow \Downarrow } + \ket{ \uparrow \downarrow\Downarrow } \bra{ \uparrow \downarrow \Uparrow })\\
&&+ \omega_{\mathrm{te}} (\ket{ \uparrow \downarrow\Uparrow } \bra{ \uparrow \downarrow \Uparrow} + \ket{ \uparrow \downarrow\Downarrow } \bra{ \uparrow \downarrow \Downarrow})
\end{eqnarray*}
where the last term corresponds to the optical transition energy $\omega_{\mathrm{te}}$.

A $\sigma_+$-polarized laser pulse couples the states $\ket{ \uparrow }$ and $\ket{ \uparrow \downarrow \Uparrow }$ \cite{atac13} while leaving $\ket{\downarrow}$ unaffected. Hence, $\ket{ \uparrow}$ is called bright and $\ket{ \downarrow}$ dark state. The duration of the laser pulse $t_{\mathrm{p}}$ is taken to be short enough so that the coherent evolution due to the four-level Hamiltonian $\hat{H}_0$ and spontaneous emission
are negligible during the pulse. We justify later that all elements of the electron density matrix $\hat{\rho}$ related to the excited states are zero at the arrival of a pulse. Then
the populations before ($\mathrm{i}$) and after ($\mathrm{f}$) a single pulse are given by \cite{supp}
\begin{equation}\label{eq:pulse}
\begin{aligned}
\rho_{\uparrow,\uparrow}^{\mathrm{f}} &= \frac{1}{2} \rho_{\uparrow,\uparrow}^{\mathrm{i}} (1+\cos\theta)\\
\rho_{\uparrow \downarrow \Uparrow,\uparrow \downarrow \Uparrow}^{\mathrm{f}} &= \frac{1}{2} \rho_{\uparrow,\uparrow}^{\mathrm{i}} (1-\cos\theta)
\end{aligned}
\end{equation}
where the pulse area $\theta = 2 \int \Omega(t)\, \mathrm dt$ determines the excited electron population and $\Omega$ is the laser Rabi frequency.

The combined coupling of the four-level system to a periodic laser pulse train and the radiation field reservoir asymptotically forces the electron into a periodic quasi-steady state. This occurs by sequential excitation of the bright state by a laser pulse, followed by spontaneous emission of a photon at a rate $\Gamma$ and the accompanying decay into one of the two ground states. If the pulse separation $\tau$ satisfies $\tau \gg \Gamma^{-1}$ the trion population decays completely within the pulse separation $\tau$, which justifies \eqref{eq:pulse}. The asymptotic bright state population
$\rho_{\uparrow,\uparrow}^{\mathrm{i},*}$ before the pulse is then given by \cite{supp}
\begin{equation*}
\rho_{\uparrow,\uparrow}^{\mathrm{i},*} = \frac{\sin^2(\omega_{\mathrm{e}} \tau/2)}{\frac{1}{2}+\sin^2(\omega_{\mathrm{e}} \tau/2)}
\end{equation*}
provided $\theta = \pi$ and $\omega_\mathrm{e} \gg \Gamma$. Similarly, the coarse-grained
spontaneous emission rate is given by
\begin{eqnarray*}
\Gamma_{\mathrm{spon}} &=& \frac{1}{\tau} \int_{0}^{\tau}
\frac{\Gamma}{2}\left[\rho_{\uparrow \downarrow \Uparrow,\uparrow \downarrow \Uparrow}(t)+\rho_{\uparrow \downarrow \Downarrow,\uparrow \downarrow \Downarrow}(t)\right]\, \mathrm dt \\
&=&\frac{1}{\tau} \rho_{\uparrow,\uparrow}^{\mathrm{i},*} =  \frac{1}{\tau} \frac{\sin^2(\omega_{\mathrm{e}} \tau/2)}{\frac{1}{2}+\sin^2(\omega_{\mathrm{e}} \tau/2)}.
\end{eqnarray*}

If the bright state population at the arrival of the pulse vanishes, so does the optical absorption and hence the coarse-grained spontaneous emission rate $\Gamma_{\mathrm{spon}}$.
This occurs exactly if the electron precession frequency satisfies the synchronization condition
\begin{equation}\label{eq:trap}
\omega_{\mathrm{e}} \in \frac{2 \pi}{\tau} \mathbb{Z}.
\end{equation}
The expansion of $\Gamma_{\mathrm{spon}}$ close to a synchronization condition \eqref{eq:trap} with $m \in \mathbb{Z}$ gives
\begin{equation*}
\Gamma_{\mathrm{spon}}= \frac{\tau}{2} \left(\omega_{\mathrm{e}}-\frac{2\pi}{\tau}m\right)^2+\mathcal{O}\left(\left(\omega_{\mathrm{e}}-\frac{2\pi}{\tau}m\right)^4\right).
\end{equation*}
The fact that $\Gamma_{\mathrm{spon}}$ depends quadratically on $\omega_{\mathrm{e}}$ is essential for the emergence of L\'evy flights.

The coupling of the central spin to the environment is given by the Fermi-contact hyperfine interaction between the QD electron and the nuclear spin environment \cite{atac13, vink09, latta09} $\hat{H}_{\mathrm{hf}} = \sum_{j=1}^N A_j \hat{S}_z \hat{I}_z^j + \sum_{j=1}^N \frac{A_j}{2} (\hat{S}_- \hat{I}_+^j + \hat{S}_+ \hat{I}_-^j)$.  The number of nuclear spins is $N\approx 10^4 - 10^6$ and $\hat{I}_z^j$ is the spin projection operator to the $z$-axis of the $j$\textsuperscript{th} nucleus for $1\le j \le N$.
The hyperfine coupling constant $A_j$ depends on the overlap of the electron wave function with the $j$\textsuperscript{th} nucleus. Equal coupling of all nuclei to the
electron leads to the homogeneous hyperfine coupling constant $A = \frac{A_{\mathrm{H}}}{N} = \frac{1}{N} \sum_{j=1}^N A_j$, where $A_{\mathrm{H}}$ is the hyperfine coupling constant of the material, assuming a single nuclear species.
The first term of $\hat{H}_{\mathrm{hf}}$, which we denote by $\hat{H}_{\mathrm{Oh}}$, describes the Overhauser shift \cite{atac13}; the electron precesses around the vector sum of the external field and the effective magnetic Overhauser field $B_{\mathrm{Oh}} = \sum_{j=1}^N A_j \langle \hat{I}_z^j \rangle/(g_{\mathrm{e}} \mu_{\mathrm{B}})$, created by the nuclear spins. The associated Overhauser precession frequency is $\omega_{\mathrm{Oh}} = \sum_{j=1}^N A_j \langle\hat{I}_z^j\rangle$. The second term of $\hat{H}_{\mathrm{hf}}$  describes nuclear spin flip-flop processes that are
suppressed at high magnetic fields due to the energy mismatch between the electron Zeeman energy ($\omega_{\mathrm{e}}$) and the average nuclear Zeeman energy ($\omega_{\mathrm{n}}$), i.e. $\omega_{\mathrm{e}} \gg \omega_{\mathrm{n}}$ \cite{atac13}.

In self-assembled QDs, large anisotropic strain gives rise to strong nuclear quadrupolar interactions where the underlying electric field gradients point primarily along the $x$-axis. This quadrupolar interaction in conjunction with the Overhauser field term constitutes the dominant mechanism for optically assisted nuclear spin flips in self-assembled QDs \cite{atac13}; we therefore neglect the flip-flop process from here on. However, we note that the flip-flop processes in $\hat{H}_{\mathrm{hf}}$, an effective non-collinear interaction of the form $\hat{H}_{\mathrm{nc}} = \sum_j A_{\mathrm{nc}}\hat{S}_z \hat{I}_x^j$ \cite{hogele12,atac13,bulutay12} or a hole-induced non-collinear process \cite{Yang2012} would yield the same dynamics. The dominant nuclear spin dynamics is described by the Hamiltonians
\begin{eqnarray*}
\hat{H}_{\mathrm{n}} &=& -\sum_{j=1}^N \omega_{\mathrm{n}}^j \hat{I}_z^j, \\
\hat{H}_{\mathrm{Oh}} &=& \sum_{j=1}^N A_j \hat{S}_z \hat{I}_z^j, \\
\hat{H}_{\mathrm{Q}} &=& \sum_{j=1}^N \frac{\omega_{\mathrm{Q}}^j}{2} \left((\hat{I}_x^j)^2 - \frac{I^j(I^j+1)}{3}\hat{\mathbb{1}}^j\right)
\end{eqnarray*}
where $\omega_{\mathrm{n}}^j$ is the Zeeman splitting of the $j$\textsuperscript{th} nucleus, and $\omega_{\mathrm{Q}}^j$ is the quadrupolar energy for $1 \le j \le N$.

The effect of the electron-nuclei interaction on the coupled system is twofold.
On the one hand, the electron dynamics is modified by the Overhauser field which changes $\omega_{\mathrm{e}}$ to the effective electron precession frequency $\omega_{\mathrm{e}}^{\mathrm{eff}} = \omega_{\mathrm{e}}+\omega_{\mathrm{Oh}}$.
The dark state is then only reached at the arrival of a pulse if synchronization \eqref{eq:trap} is fulfilled for the effective precession frequency
\begin{equation}
\omega_{\mathrm{e}}^{\mathrm{eff}} \in \frac{2\pi}{\tau}\mathbb{Z}. \label{eq:trap:eff}
\end{equation}
The modified electron dynamics results in a coarse-grained spontaneous emission rate $\Gamma_{\mathrm{spon}}$ that depends on $\omega_{\mathrm{e}}^{\mathrm{eff}}$ and thus on the nuclear Overhauser field $B_{\mathrm{Oh}}$, and vanishes if the synchronization condition is satisfied.

\begin{figure}
 \includegraphics[width=8.6cm]{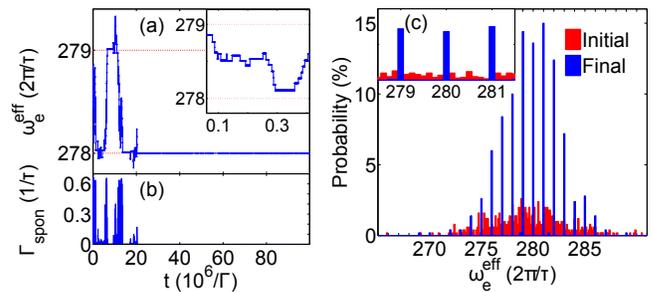}
 \caption{
(a) The random walk of the effective electron precession frequency (blue) spends most of the time close to traps (red), indicating a L\'evy distribution for long waiting times. The inset resolves the random walk between the traps.
(b) $\Gamma_{\mathrm{spon}}$ gives the spontaneous emission rate of the corresponding random walk in (a).
(c) Exposing an ensemble of 500 QDs with normally distributed initial nuclear polarization ($\langle I_z \rangle = 5$, $\sigma = \sqrt{N}$) (red) to the pulsed laser train for $10^8\tau = 10^9/\Gamma$ leads to a final (blue) distribution of the effective electron precession frequency that is heavily peaked at the different traps (inset).
 For all simulations we used the parameters $\tau = 10/\Gamma, \omega_{\mathrm{e}} = 280 \frac{2\pi}{\tau}, \omega_{\mathrm{n}} = 0.18 \Gamma, N = 2000, \omega_{\mathrm{Q}} = 0.013 \Gamma, A_{\mathrm{H}} = 100\Gamma$, and we assumed inhomogeneous hyperfine coupling constants. 
 \label{fig:rw}}
\end{figure}

On the other hand, the nuclear dynamics is modified by optical excitation of the electron, which results in nuclear spin flip assisted spontaneous emission and leads to a slow time evolution of the nuclear Overhauser field. To describe such processes, we apply a Schrieffer-Wolff transformation \cite{shavitt80, issler10} to eliminate the non-energy conserving terms of $\hat{H}_{\mathrm{Q}}$.
We find second-order nuclear spin flip processes where the additional energy is supplied by a spontaneously emitted photon (FIG. \ref{fig:coupling} (b)) occurring with a rate \cite{supp}
\begin{eqnarray*}
R_{\mathrm{nuc}} &=& \epsilon^2 \Gamma_{\mathrm{spon}} \\
        &=& \epsilon^2 \frac{1}{\tau} \frac{\sin^2(\omega_{\mathrm{e}}^{\mathrm{eff}} \tau/2)}{\frac{1}{2}+\sin^2(\omega_{\mathrm{e}}^{\mathrm{eff}} \tau/2)}\\
        &=& \epsilon^2 \frac{\tau}{2} \left(\omega_{\mathrm{e}}^{\mathrm{eff}} - \frac{2\pi}{\tau}m\right)^2+\mathcal{O}\left(\left(\omega_{\mathrm{e}}^{\mathrm{eff}}-\frac{2\pi}{\tau}m\right)^4\right)
\end{eqnarray*}
where $\epsilon = \sqrt{N} \frac{A\omega_{\mathrm{Q}}}{32\omega_{\mathrm{n}}^2}$ under the assumptions of homogeneous hyperfine and quadrupolar coupling, $\omega_{\mathrm{n}}\gg A$ and $I_z^j = \pm \frac{1}{2}$ for simplicity \cite{spin_1_2}.
The rate of optically-assisted nuclear spin flips $R_{\mathrm{nuc}}$ is drastically reduced
in the vicinity of a synchronization condition \eqref{eq:trap:eff}. The signatures of this reduction can be observed in a single realization of a random walk of $\omega_{\mathrm{e}}^{\mathrm{eff}}$ ($I_z$): the random walk freezes near a synchronization condition which therefore constitutes a trap (red lines in FIG. \ref{fig:rw} (a)).
An unsynchronized electron is optically excited and induces an unbiased random walk of the nuclear polarization $I_z = \sum_{j=1}^{N} \langle I_z^j\rangle$ until $\omega_{\mathrm{e}}^{\mathrm{eff}}$ is synchronized. In the absence of other interactions, the time spent near such a trap can be on the order of the measurement time. We also simulated the time evolution of the distribution of $\omega_{\mathrm{e}}^{\mathrm{eff}}$ starting from a Gaussian distribution of $\{I_z^j\}_{j=1}^N$ (red histogram in FIG. \ref{fig:rw} (c)). We find that the probability of finding $\omega_{\mathrm{e}}^{\mathrm{eff}}$ within a single nuclear spin flip from a trap \eqref{eq:trap:eff} exceeds $99 \%$. These simulations are in excellent qualitative agreement with experiments carried out on an ensemble of single negatively-charged self-assembled QDs \cite{bayer07}.

To quantify the freezing of a random walk of $\omega_{\mathrm{e}}^{\mathrm{eff}}$ in the vicinity of a trap, we introduce the probability distribution of the waiting time between two successive nuclear spin flip events $P_{\mathrm{nuc}}(\Delta t_{\mathrm{nuc}})$ \cite{bardou94}. Clearly, long waiting times occur for  $\omega_{\mathrm{e}}^{\mathrm{eff}}$ close to a trap since $\Delta t_{\mathrm{nuc}} \sim R_{{\mathrm{nuc}}}^{-1} \propto (\omega_{\mathrm{e}}^{\mathrm{eff}} - \frac{2\pi}{\tau}m)^{-2}$. This relationship of $\Delta t_{\mathrm{nuc}}$ to $\omega_{\mathrm{e}}^{\mathrm{eff}}$ allows us to write $P_{{\mathrm{nuc}}}(\Delta t_{\mathrm{nuc}}) \mathrm d\Delta t_{\mathrm{nuc}} = \Pi_{{\mathrm{nuc}}}(\omega_{\mathrm{e}}^{\mathrm{eff}}) \mathrm d \omega_{\mathrm{e}}^{\mathrm{eff}}$, where the probability to reach an effective precession frequency $\omega_{\mathrm{e}}^{\mathrm{eff}}$ in the trap $\Pi_{{\mathrm{nuc}}}(\omega_{\mathrm{e}}^{\mathrm{eff}})$ is assumed to be uniform. Estimating the change in $\Delta t_{\mathrm{nuc}}$ induced by an infinitesimal change in $\omega_{\mathrm{e}}^{\mathrm{eff}}$ via $\frac{\mathrm d\Delta t_{\mathrm{nuc}}}{\mathrm d\omega_{\mathrm{e}}^{\mathrm{eff}}} \propto (\omega_{\mathrm{e}}^{\mathrm{eff}}-\frac{2 \pi}{\tau}m)^{-3} \propto \Delta t_{\mathrm{nuc}}^{3/2}$, results in the power-law distribution
\begin{equation}\label{eq:waiting}
P_{{\mathrm{nuc}}}(\Delta t_{\mathrm{nuc}}) \propto \Delta t_{\mathrm{nuc}}^{-3/2}.
\end{equation}
This expression is confirmed by a more rigorous derivation \cite{levyBook} or \cite{supp} and is valid for $\Delta t_{\mathrm{nuc}} \gg \left(\left.R_{\mathrm{nuc}}\right \vert_{\omega_{\mathrm{e}}^{\mathrm{eff}}=\frac{2\pi}{\tau}m\pm A}\right)^{-1}$ \cite{t_1,supp}. The expectation value of this distribution does not exist. The L\'evy behavior of the nuclear spin flip waiting time is thus based on the quadratic dependence of $\Gamma_{\mathrm{spon}}$ and $R_{{\mathrm{nuc}}}$ on $\omega_{\mathrm{e}}^{\mathrm{eff}}$.

The physical observable in an experiment is the waiting time $\Delta t_{\mathrm{spon}}$ between two successive photon emission events, irrespective of whether the photon emission is accompanied by a nuclear spin flip or not. The rate of photon emission events is given by $\Gamma_{\mathrm{spon}}$ and an analogous derivation yields the waiting time distribution of successive photon emission events \cite{supp}
\begin{equation}\label{eq:waiting:pure}
P_{\mathrm{spon}}(\Delta t_{\mathrm{spon}}) \propto \Delta t_{\mathrm{spon}}^{-3/2},
\end{equation}
for $\Delta t_{\mathrm{spon}} \gg \epsilon^2 \left(\left.R_{\mathrm{nuc}}\right \vert_{\omega_{\mathrm{e}}^{\mathrm{eff}}=\frac{2\pi}{\tau}m\pm A}\right)^{-1}$. As a consequence of the proportionality of the rate of nuclear spin flip events $R_{\mathrm{nuc}}$ to spontaneous emission events $\Gamma_{\mathrm{spon}}$, the power-law exponents of the two waiting time distributions $P_{\mathrm{nuc}}$ and $P_{\mathrm{spon}}$ are identical. Since the rate of photon detection events is enhanced by a factor $\epsilon^{-2}$ compared to the nuclear spin flip rate, it is possible to monitor the L\'evy flight behavior of nuclear trapping times in real-time even with modest photon detection efficiencies.

\begin{figure}
\centering
 \includegraphics[width=8.6cm]{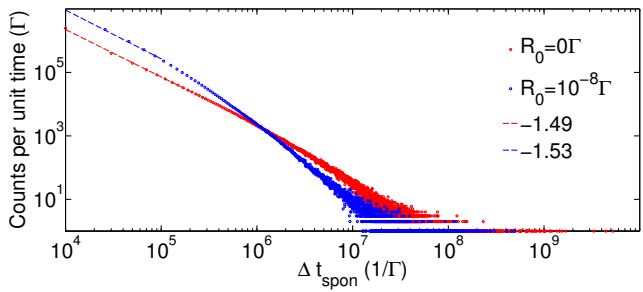}
 \caption{The optical waiting time distribution for $R_0 = 0\Gamma$ (red) is extracted from 100 time traces. The corresponding power-law exponent in this simulation is $-1.49$, consistent with the theoretical prediction of $-3/2$. For $R_0^{-1} = 10^{8}/\Gamma$ (blue) we find a regime where a power-law with exponent $-1.53$ is observed, followed by an exponential decay. All other parameters are the same as in FIG. \ref{fig:rw}.
 \label{fig:offset}}
\end{figure}

To numerically verify the power-law behavior of the waiting time distribution, we averaged over $100$ Monte-Carlo simulations of the type shown in FIG. \ref{fig:rw} (a) and extracted a power-law exponent $\eta = -1.49$
(red dotted line in FIG. \ref{fig:offset}) which is in excellent agreement with the predicted distribution of optical waiting times \eqref{eq:waiting:pure}.

Experimentally observable waiting times are limited due to additional interactions that have been neglected so far. To investigate their impact on the probability of nuclear trapping times, we assume that they can be modeled as an additional constant nuclear spin flip rate $R_0$ such that the overall nuclear spin flip rate becomes
\begin{equation*}
\tilde{R}_{\mathrm{nuc}} = R_{\mathrm{nuc}}+R_0.
\end{equation*}
The constant rate $R_0$ prevents diverging average trapping times since it effectively limits the trapping times to $R_0^{-1}$. This scenario corresponds to a truncated L\'evy flight \cite{mantegna94}. We derived the modified distribution of nuclear trapping times \cite{supp}
\begin{equation*}
\tilde{P}_{\mathrm{nuc}}(\Delta t_{\mathrm{nuc}}) \propto \Delta t_{\mathrm{nuc}}^{-3/2}(1 + 2R_0\Delta t_{\mathrm{nuc}}) \mathrm{e}^{-R_0 \Delta t_{\mathrm{nuc}}}.
\end{equation*}
If $R_0 = 0$ then expression \eqref{eq:waiting} is recovered, but if $\Delta t_{\mathrm{nuc}} \gg R_0^{-1}$ the exponential term dominates and leads to an exponential decay. However, in an intermediate regime, where the background rate $R_0$ is much smaller than $R_{\mathrm{nuc}}$ at a single spin flip away from synchronization, i.e. $R_0 \ll \left.R_{\mathrm{nuc}}\right \vert_{\omega_{\mathrm{e}}^{\mathrm{eff}}=m\frac{2\pi}{\tau} + A}$, a power-law regime is observed before the exponential cutoff.
The optical waiting time distribution $\tilde{P}_{\mathrm{spon}}$ with constant offset $R_0$ is again closely related to the nuclear trapping time distribution $\tilde{P}_{\mathrm{nuc}}$. A model simulation including a constant offset $R_0 \neq 0\Gamma$ shows a power law distribution with coefficient $-1.53$ (blue dotted line in FIG. \ref{fig:offset}) for intermediate waiting times and an exponential decay for large waiting times.

The physical origin of $R_0$ can be classified as system-specific or intrinsic to the scheme. In the case of an electron localized in a self-assembled QD the coupling of the electron spin to phonons \cite{erlingsson01} or to a Fermionic reservoir \cite{dreiser08} could limit the trapping times. In addition, finite pulse duration and imperfect laser polarization lead to a finite decay rate out of the electronic dark state.
These effects can be suppressed by proper engineering. In contrast, the scheme can be intrinsically limited by the interaction that causes the optically assisted random walk, which is the quadrupolar interaction $\hat{H}_{Q}$ for self-assembled QDs. In the time windows that separate the successive laser pulses, $\hat{H}_{Q}$ leads to a coherent precession between different $\hat{I}_z$ eigenstates; the precession frequency is dominated by the energy detuning $\sim \omega_{\mathrm{n}}$. In the absence of a measurement interaction interrupting the coherent evolution no actual nuclear spin flip can occur due to the energy mismatch of the coherently coupled nuclear states.
However, the laser pulse terminating the period of coherent evolution under $\hat{H}_{\mathrm{Q}}$ effects such a measurement and could lead to a nuclear spin flip.
Consequently, the rate at which these intrinsic nuclear spin flips occur depends on the bright population in a different nuclear spin manifold at the arrival of a pulse. Similar intrinsically limiting mechanisms appear for systems with dominant non-collinear or hyperfine flip-flop interactions, where the latter is discussed in \cite{economou14}.

The intrinsic contribution to $R_0$ can be minimized by tuning the magnetic field $B$ and pulse repetition period $\tau$. The coherent evolution of the nuclear spins is dominated by the nuclear Zeeman frequency $\omega_{\mathrm{n}}$ and can be reduced if the nuclei are also synchronized \cite{supp}
\begin{equation}
\omega_{\mathrm{n}} \in \frac{2\pi}{\tau}\mathbb{Z}.
\end{equation}
In a system with several nuclear species it could be challenging to synchronize all nuclei for an experimentally achievable $B$ and $\tau$. Adding an echo pulse on the electron would add flexibility to synchronize the nuclei and the electron separately. In this case, it is important to displace the echo pulse by a finite $\Delta t_{\mathrm{echo}}$ from $\tau/2$ to ensure that the electron dynamics responsible for the optically assisted nuclear spin flips at a rate $R_{\mathrm{nuc}}$ is not completely reversed. The displaced echo pulses define a new synchronization condition for the electron which is given by \cite{supp}
\begin{equation*}
\omega_{\mathrm{e}}^{\mathrm{eff}} \in \frac{\pi}{\Delta t_{\mathrm{echo}}}\mathbb{Z}.
\end{equation*}
This procedure adds an additional parameter $\Delta t_{\mathrm{echo}}$ that controls the electron dynamics, while $\tau$ can be extended to accommodate better synchronization for multiple nuclear species without affecting the electron dynamics.

In summary, we have demonstrated that the polarization of the spin environment of a central spin system that is driven by a resonant periodic pulse train exhibits L\'evy flights. Thanks to the decoupling of the central spin and the environment and the proportionality of the environmental spin flip and spontaneous emission rates, our calculations indicate that the signatures of L\'evy flights can be observed by measuring the waiting time distribution of the emitted photons. While we illustrated the scheme with self-assembled quantum dots, we expect our findings to be relevant for a broad class of solid-state emitters including NV centers. More generally, our results show that the previously established connection between L\'evy flights and dark states \cite{bardou94,issler10} can be extended to systems with explicit time-dependence.

\bibliography{LevyFlights-MainPaper}

\begin{thebibliography}{23}%
\makeatletter
\providecommand \@ifxundefined [1]{%
 \@ifx{#1\undefined}
}%
\providecommand \@ifnum [1]{%
 \ifnum #1\expandafter \@firstoftwo
 \else \expandafter \@secondoftwo
 \fi
}%
\providecommand \@ifx [1]{%
 \ifx #1\expandafter \@firstoftwo
 \else \expandafter \@secondoftwo
 \fi
}%
\providecommand \natexlab [1]{#1}%
\providecommand \enquote  [1]{``#1''}%
\providecommand \bibnamefont  [1]{#1}%
\providecommand \bibfnamefont [1]{#1}%
\providecommand \citenamefont [1]{#1}%
\providecommand \href@noop [0]{\@secondoftwo}%
\providecommand \href [0]{\begingroup \@sanitize@url \@href}%
\providecommand \@href[1]{\@@startlink{#1}\@@href}%
\providecommand \@@href[1]{\endgroup#1\@@endlink}%
\providecommand \@sanitize@url [0]{\catcode `\\12\catcode `\$12\catcode
  `\&12\catcode `\#12\catcode `\^12\catcode `\_12\catcode `\%12\relax}%
\providecommand \@@startlink[1]{}%
\providecommand \@@endlink[0]{}%
\providecommand \url  [0]{\begingroup\@sanitize@url \@url }%
\providecommand \@url [1]{\endgroup\@href {#1}{\urlprefix }}%
\providecommand \urlprefix  [0]{URL }%
\providecommand \Eprint [0]{\href }%
\providecommand \doibase [0]{http://dx.doi.org/}%
\providecommand \selectlanguage [0]{\@gobble}%
\providecommand \bibinfo  [0]{\@secondoftwo}%
\providecommand \bibfield  [0]{\@secondoftwo}%
\providecommand \translation [1]{[#1]}%
\providecommand \BibitemOpen [0]{}%
\providecommand \bibitemStop [0]{}%
\providecommand \bibitemNoStop [0]{.\EOS\space}%
\providecommand \EOS [0]{\spacefactor3000\relax}%
\providecommand \BibitemShut  [1]{\csname bibitem#1\endcsname}%
\let\auto@bib@innerbib\@empty
\bibitem [{\citenamefont {Shlesinger}\ \emph {et~al.}(1993)\citenamefont
  {Shlesinger}, \citenamefont {Zaslavsky},\ and\ \citenamefont
  {Klafter}}]{shlesinger93}%
  \BibitemOpen
  \bibfield  {author} {\bibinfo {author} {\bibfnamefont {M.~F.}\ \bibnamefont
  {Shlesinger}}, \bibinfo {author} {\bibfnamefont {G.~M.}\ \bibnamefont
  {Zaslavsky}}, \ and\ \bibinfo {author} {\bibfnamefont {J.}~\bibnamefont
  {Klafter}},\ }\href {\doibase 10.1038/363031a0} {\bibfield  {journal}
  {\bibinfo  {journal} {Nature}\ }\textbf {\bibinfo {volume} {363}},\ \bibinfo
  {pages} {31} (\bibinfo {year} {1993})}\BibitemShut {NoStop}%
\bibitem [{\citenamefont {Barthelemy}\ \emph {et~al.}(2008)\citenamefont
  {Barthelemy}, \citenamefont {Bertolotti},\ and\ \citenamefont
  {Wiersma}}]{barthelemy08}%
  \BibitemOpen
  \bibfield  {author} {\bibinfo {author} {\bibfnamefont {P.}~\bibnamefont
  {Barthelemy}}, \bibinfo {author} {\bibfnamefont {J.}~\bibnamefont
  {Bertolotti}}, \ and\ \bibinfo {author} {\bibfnamefont {D.~S.}\ \bibnamefont
  {Wiersma}},\ }\href {\doibase 10.1038/nature06948} {\bibfield  {journal}
  {\bibinfo  {journal} {Nature}\ }\textbf {\bibinfo {volume} {453}},\ \bibinfo
  {pages} {495} (\bibinfo {year} {2008})}\BibitemShut {NoStop}%
\bibitem [{\citenamefont {Bardou}\ \emph {et~al.}(1994)\citenamefont {Bardou},
  \citenamefont {Bouchaud}, \citenamefont {Emile}, \citenamefont {Aspect},\
  and\ \citenamefont {Cohen-Tannoudji}}]{bardou94}%
  \BibitemOpen
  \bibfield  {author} {\bibinfo {author} {\bibfnamefont {F.}~\bibnamefont
  {Bardou}}, \bibinfo {author} {\bibfnamefont {J.}~\bibnamefont {Bouchaud}},
  \bibinfo {author} {\bibfnamefont {O.}~\bibnamefont {Emile}}, \bibinfo
  {author} {\bibfnamefont {A.}~\bibnamefont {Aspect}}, \ and\ \bibinfo {author}
  {\bibfnamefont {C.}~\bibnamefont {Cohen-Tannoudji}},\ }\href
  {https://urldefense.proofpoint.com/v2/url?u=http-3A__www.ncbi.nlm.nih.gov_pubmed_10056085&d=AwIG-g&c=-dg2m7zWuuDZ0MUcV7Sdqw&r=wjB15AxuD9LZg8OWh3q44sP_btMUEdB-Zf79J-wYH_I&m=eGjAXwXFgPKHeLjt85VjZIZzpkxu3N3_2_WSbDCvs6c&s=Wz822ViN9krI6xh_hrv4VrzGrx4j9mBXl-84z_H4oF8&e=}
  {\bibfield  {journal} {\bibinfo  {journal} {Physical Review Letters}\
  }\textbf {\bibinfo {volume} {72}},\ \bibinfo {pages} {203} (\bibinfo {year}
  {1994})}\BibitemShut {NoStop}%
\bibitem [{\citenamefont {Greilich}\ \emph {et~al.}(2006)\citenamefont
  {Greilich}, \citenamefont {Oulton}, \citenamefont {Zhukov}, \citenamefont
  {Yugova}, \citenamefont {Yakovlev}, \citenamefont {Bayer}, \citenamefont
  {Shabaev}, \citenamefont {Efros}, \citenamefont {Merkulov}, \citenamefont
  {Stavarache}, \citenamefont {Reuter},\ and\ \citenamefont
  {Wieck}}]{greilich06}%
  \BibitemOpen
  \bibfield  {author} {\bibinfo {author} {\bibfnamefont {A.}~\bibnamefont
  {Greilich}}, \bibinfo {author} {\bibfnamefont {R.}~\bibnamefont {Oulton}},
  \bibinfo {author} {\bibfnamefont {E.}~\bibnamefont {Zhukov}}, \bibinfo
  {author} {\bibfnamefont {I.}~\bibnamefont {Yugova}}, \bibinfo {author}
  {\bibfnamefont {D.}~\bibnamefont {Yakovlev}}, \bibinfo {author}
  {\bibfnamefont {M.}~\bibnamefont {Bayer}}, \bibinfo {author} {\bibfnamefont
  {A.}~\bibnamefont {Shabaev}}, \bibinfo {author} {\bibfnamefont
  {A.}~\bibnamefont {Efros}}, \bibinfo {author} {\bibfnamefont
  {I.}~\bibnamefont {Merkulov}}, \bibinfo {author} {\bibfnamefont
  {V.}~\bibnamefont {Stavarache}}, \bibinfo {author} {\bibfnamefont
  {D.}~\bibnamefont {Reuter}}, \ and\ \bibinfo {author} {\bibfnamefont
  {A.}~\bibnamefont {Wieck}},\ }\href {\doibase 10.1103/PhysRevLett.96.227401}
  {\bibfield  {journal} {\bibinfo  {journal} {Physical Review Letters}\
  }\textbf {\bibinfo {volume} {96}},\ \bibinfo {pages} {227401} (\bibinfo
  {year} {2006})}\BibitemShut {NoStop}%
\bibitem [{\citenamefont {Greilich}\ \emph {et~al.}(2007)\citenamefont
  {Greilich}, \citenamefont {Shabaev}, \citenamefont {Yakovlev}, \citenamefont
  {Efros}, \citenamefont {Yugova}, \citenamefont {Reuter}, \citenamefont
  {Wieck},\ and\ \citenamefont {Bayer}}]{bayer07}%
  \BibitemOpen
  \bibfield  {author} {\bibinfo {author} {\bibfnamefont {a.}~\bibnamefont
  {Greilich}}, \bibinfo {author} {\bibfnamefont {A.}~\bibnamefont {Shabaev}},
  \bibinfo {author} {\bibfnamefont {D.~R.}\ \bibnamefont {Yakovlev}}, \bibinfo
  {author} {\bibfnamefont {A.~L.}\ \bibnamefont {Efros}}, \bibinfo {author}
  {\bibfnamefont {I.~a.}\ \bibnamefont {Yugova}}, \bibinfo {author}
  {\bibfnamefont {D.}~\bibnamefont {Reuter}}, \bibinfo {author} {\bibfnamefont
  {a.~D.}\ \bibnamefont {Wieck}}, \ and\ \bibinfo {author} {\bibfnamefont
  {M.}~\bibnamefont {Bayer}},\ }\href {\doibase 10.1126/science.1146850}
  {\bibfield  {journal} {\bibinfo  {journal} {Science (New York, N.Y.)}\
  }\textbf {\bibinfo {volume} {317}},\ \bibinfo {pages} {1896} (\bibinfo {year}
  {2007})}\BibitemShut {NoStop}%
\bibitem [{\citenamefont {Barnes}\ and\ \citenamefont
  {Economou}(2011)}]{barnes11}%
  \BibitemOpen
  \bibfield  {author} {\bibinfo {author} {\bibfnamefont {E.}~\bibnamefont
  {Barnes}}\ and\ \bibinfo {author} {\bibfnamefont {S.~E.}\ \bibnamefont
  {Economou}},\ }\href {\doibase 10.1103/PhysRevLett.107.047601} {\bibfield
  {journal} {\bibinfo  {journal} {Physical Review Letters}\ }\textbf {\bibinfo
  {volume} {107}},\ \bibinfo {pages} {047601} (\bibinfo {year}
  {2011})}\BibitemShut {NoStop}%
\bibitem [{\citenamefont {Economou}\ and\ \citenamefont
  {Barnes}(2014)}]{economou14}%
  \BibitemOpen
  \bibfield  {author} {\bibinfo {author} {\bibfnamefont {S.~E.}\ \bibnamefont
  {Economou}}\ and\ \bibinfo {author} {\bibfnamefont {E.}~\bibnamefont
  {Barnes}},\ }\href {\doibase 10.1103/PhysRevB.89.165301} {\bibfield
  {journal} {\bibinfo  {journal} {Physical Review B}\ }\textbf {\bibinfo
  {volume} {89}},\ \bibinfo {pages} {165301} (\bibinfo {year}
  {2014})}\BibitemShut {NoStop}%
\bibitem [{\citenamefont {Yugova}\ \emph {et~al.}(2009)\citenamefont {Yugova},
  \citenamefont {Glazov}, \citenamefont {Ivchenko},\ and\ \citenamefont
  {Efros}}]{yugova09}%
  \BibitemOpen
  \bibfield  {author} {\bibinfo {author} {\bibfnamefont {I.~A.}\ \bibnamefont
  {Yugova}}, \bibinfo {author} {\bibfnamefont {M.~M.}\ \bibnamefont {Glazov}},
  \bibinfo {author} {\bibfnamefont {E.~L.}\ \bibnamefont {Ivchenko}}, \ and\
  \bibinfo {author} {\bibfnamefont {A.~L.}\ \bibnamefont {Efros}},\ }\href
  {\doibase 10.1103/PhysRevB.80.104436} {\bibfield  {journal} {\bibinfo
  {journal} {Physical Review B}\ }\textbf {\bibinfo {volume} {80}},\ \bibinfo
  {pages} {104436} (\bibinfo {year} {2009})},\ \Eprint
  {http://arxiv.org/abs/0905.1254} {0905.1254} \BibitemShut {NoStop}%
\bibitem [{\citenamefont {Urbaszek}\ \emph {et~al.}(2013)\citenamefont
  {Urbaszek}, \citenamefont {Marie}, \citenamefont {Amand}, \citenamefont
  {Krebs}, \citenamefont {Voisin}, \citenamefont {Maletinsky}, \citenamefont
  {H\"{o}gele},\ and\ \citenamefont {Imamoglu}}]{atac13}%
  \BibitemOpen
  \bibfield  {author} {\bibinfo {author} {\bibfnamefont {B.}~\bibnamefont
  {Urbaszek}}, \bibinfo {author} {\bibfnamefont {X.}~\bibnamefont {Marie}},
  \bibinfo {author} {\bibfnamefont {T.}~\bibnamefont {Amand}}, \bibinfo
  {author} {\bibfnamefont {O.}~\bibnamefont {Krebs}}, \bibinfo {author}
  {\bibfnamefont {P.}~\bibnamefont {Voisin}}, \bibinfo {author} {\bibfnamefont
  {P.}~\bibnamefont {Maletinsky}}, \bibinfo {author} {\bibfnamefont
  {A.}~\bibnamefont {H\"{o}gele}}, \ and\ \bibinfo {author} {\bibfnamefont
  {A.}~\bibnamefont {Imamoglu}},\ }\href {\doibase 10.1103/RevModPhys.85.79}
  {\bibfield  {journal} {\bibinfo  {journal} {Reviews of Modern Physics}\
  }\textbf {\bibinfo {volume} {85}},\ \bibinfo {pages} {79} (\bibinfo {year}
  {2013})}\BibitemShut {NoStop}%
\bibitem [{\citenamefont {Issler}\ \emph {et~al.}()\citenamefont {Issler},
  \citenamefont {H\"{o}ller},\ and\ \citenamefont {Imamoglu}}]{supp}%
  \BibitemOpen
  \bibfield  {author} {\bibinfo {author} {\bibfnamefont {M.}~\bibnamefont
  {Issler}}, \bibinfo {author} {\bibfnamefont {J.}~\bibnamefont {H\"{o}ller}},
  \ and\ \bibinfo {author} {\bibfnamefont {A.}~\bibnamefont {Imamoglu}},\
  }\href@noop {} {}\bibinfo {note} {Supplementary material in
  preparation}\BibitemShut {NoStop}%
\bibitem [{\citenamefont {Vink}\ \emph {et~al.}(2009)\citenamefont {Vink},
  \citenamefont {Nowack}, \citenamefont {Koppens}, \citenamefont {Danon},
  \citenamefont {Nazarov},\ and\ \citenamefont {Vandersypen}}]{vink09}%
  \BibitemOpen
  \bibfield  {author} {\bibinfo {author} {\bibfnamefont {I.~T.}\ \bibnamefont
  {Vink}}, \bibinfo {author} {\bibfnamefont {K.~C.}\ \bibnamefont {Nowack}},
  \bibinfo {author} {\bibfnamefont {F.~H.~L.}\ \bibnamefont {Koppens}},
  \bibinfo {author} {\bibfnamefont {J.}~\bibnamefont {Danon}}, \bibinfo
  {author} {\bibfnamefont {Y.~V.}\ \bibnamefont {Nazarov}}, \ and\ \bibinfo
  {author} {\bibfnamefont {L.~M.~K.}\ \bibnamefont {Vandersypen}},\ }\href
  {\doibase 10.1038/nphys1366} {\bibfield  {journal} {\bibinfo  {journal}
  {Nature Physics}\ }\textbf {\bibinfo {volume} {5}},\ \bibinfo {pages} {764}
  (\bibinfo {year} {2009})}\BibitemShut {NoStop}%
\bibitem [{\citenamefont {Latta}\ \emph {et~al.}(2009)\citenamefont {Latta},
  \citenamefont {H\"{o}gele}, \citenamefont {Zhao}, \citenamefont {Vamivakas},
  \citenamefont {Maletinsky}, \citenamefont {Kroner}, \citenamefont {Dreiser},
  \citenamefont {Carusotto}, \citenamefont {Badolato}, \citenamefont {Schuh},
  \citenamefont {Wegscheider}, \citenamefont {Atat\"{u}re},\ and\ \citenamefont
  {Imamoglu}}]{latta09}%
  \BibitemOpen
  \bibfield  {author} {\bibinfo {author} {\bibfnamefont {C.}~\bibnamefont
  {Latta}}, \bibinfo {author} {\bibfnamefont {A.}~\bibnamefont {H\"{o}gele}},
  \bibinfo {author} {\bibfnamefont {Y.}~\bibnamefont {Zhao}}, \bibinfo {author}
  {\bibfnamefont {A.~N.}\ \bibnamefont {Vamivakas}}, \bibinfo {author}
  {\bibfnamefont {P.}~\bibnamefont {Maletinsky}}, \bibinfo {author}
  {\bibfnamefont {M.}~\bibnamefont {Kroner}}, \bibinfo {author} {\bibfnamefont
  {J.}~\bibnamefont {Dreiser}}, \bibinfo {author} {\bibfnamefont
  {I.}~\bibnamefont {Carusotto}}, \bibinfo {author} {\bibfnamefont
  {A.}~\bibnamefont {Badolato}}, \bibinfo {author} {\bibfnamefont
  {D.}~\bibnamefont {Schuh}}, \bibinfo {author} {\bibfnamefont
  {W.}~\bibnamefont {Wegscheider}}, \bibinfo {author} {\bibfnamefont
  {M.}~\bibnamefont {Atat\"{u}re}}, \ and\ \bibinfo {author} {\bibfnamefont
  {A.}~\bibnamefont {Imamoglu}},\ }\href {\doibase 10.1038/nphys1363}
  {\bibfield  {journal} {\bibinfo  {journal} {Nature Physics}\ }\textbf
  {\bibinfo {volume} {5}},\ \bibinfo {pages} {758} (\bibinfo {year}
  {2009})}\BibitemShut {NoStop}%
\bibitem [{\citenamefont {H\"{o}gele}\ \emph {et~al.}(2012)\citenamefont
  {H\"{o}gele}, \citenamefont {Kroner}, \citenamefont {Latta}, \citenamefont
  {Claassen}, \citenamefont {Carusotto}, \citenamefont {Bulutay},\ and\
  \citenamefont {Imamoglu}}]{hogele12}%
  \BibitemOpen
  \bibfield  {author} {\bibinfo {author} {\bibfnamefont {A.}~\bibnamefont
  {H\"{o}gele}}, \bibinfo {author} {\bibfnamefont {M.}~\bibnamefont {Kroner}},
  \bibinfo {author} {\bibfnamefont {C.}~\bibnamefont {Latta}}, \bibinfo
  {author} {\bibfnamefont {M.}~\bibnamefont {Claassen}}, \bibinfo {author}
  {\bibfnamefont {I.}~\bibnamefont {Carusotto}}, \bibinfo {author}
  {\bibfnamefont {C.}~\bibnamefont {Bulutay}}, \ and\ \bibinfo {author}
  {\bibfnamefont {A.}~\bibnamefont {Imamoglu}},\ }\href {\doibase
  10.1103/PhysRevLett.108.197403} {\bibfield  {journal} {\bibinfo  {journal}
  {Physical Review Letters}\ }\textbf {\bibinfo {volume} {108}},\ \bibinfo
  {pages} {197403} (\bibinfo {year} {2012})}\BibitemShut {NoStop}%
\bibitem [{\citenamefont {Bulutay}(2012)}]{bulutay12}%
  \BibitemOpen
  \bibfield  {author} {\bibinfo {author} {\bibfnamefont {C.}~\bibnamefont
  {Bulutay}},\ }\href {\doibase 10.1103/PhysRevB.85.115313} {\bibfield
  {journal} {\bibinfo  {journal} {Physical Review B}\ }\textbf {\bibinfo
  {volume} {85}},\ \bibinfo {pages} {115313} (\bibinfo {year}
  {2012})}\BibitemShut {NoStop}%
\bibitem [{\citenamefont {Yang}\ and\ \citenamefont {Sham}(2012)}]{Yang2012}%
  \BibitemOpen
  \bibfield  {author} {\bibinfo {author} {\bibfnamefont {W.}~\bibnamefont
  {Yang}}\ and\ \bibinfo {author} {\bibfnamefont {L.~J.}\ \bibnamefont
  {Sham}},\ }\href
  {https://urldefense.proofpoint.com/v2/url?u=http-3A__link.aps.org_doi_10.1103_PhysRevB.85.235319&d=AwIG-g&c=-dg2m7zWuuDZ0MUcV7Sdqw&r=wjB15AxuD9LZg8OWh3q44sP_btMUEdB-Zf79J-wYH_I&m=eGjAXwXFgPKHeLjt85VjZIZzpkxu3N3_2_WSbDCvs6c&s=eI5K-iF13bfyMG82iYJfGWXNQeBa4HqMuG7oP1X4JHM&e=}
  {\bibfield  {journal} {\bibinfo  {journal} {Physical Review B}\ }\textbf
  {\bibinfo {volume} {85}},\ \bibinfo {pages} {235319} (\bibinfo {year}
  {2012})}\BibitemShut {NoStop}%
\bibitem [{\citenamefont {Shavitt}\ and\ \citenamefont
  {Redmon}(1980)}]{shavitt80}%
  \BibitemOpen
  \bibfield  {author} {\bibinfo {author} {\bibfnamefont {I.}~\bibnamefont
  {Shavitt}}\ and\ \bibinfo {author} {\bibfnamefont {L.~T.}\ \bibnamefont
  {Redmon}},\ }\href {\doibase 10.1063/1.440050} {\bibfield  {journal}
  {\bibinfo  {journal} {The Journal of Chemical Physics}\ }\textbf {\bibinfo
  {volume} {73}},\ \bibinfo {pages} {5711} (\bibinfo {year}
  {1980})}\BibitemShut {NoStop}%
\bibitem [{\citenamefont {Issler}\ \emph {et~al.}(2010)\citenamefont {Issler},
  \citenamefont {Kessler}, \citenamefont {Giedke}, \citenamefont {Yelin},
  \citenamefont {Cirac}, \citenamefont {Lukin},\ and\ \citenamefont
  {Imamoglu}}]{issler10}%
  \BibitemOpen
  \bibfield  {author} {\bibinfo {author} {\bibfnamefont {M.}~\bibnamefont
  {Issler}}, \bibinfo {author} {\bibfnamefont {E.}~\bibnamefont {Kessler}},
  \bibinfo {author} {\bibfnamefont {G.}~\bibnamefont {Giedke}}, \bibinfo
  {author} {\bibfnamefont {S.}~\bibnamefont {Yelin}}, \bibinfo {author}
  {\bibfnamefont {I.}~\bibnamefont {Cirac}}, \bibinfo {author} {\bibfnamefont
  {M.}~\bibnamefont {Lukin}}, \ and\ \bibinfo {author} {\bibfnamefont
  {A.}~\bibnamefont {Imamoglu}},\ }\href {\doibase
  10.1103/PhysRevLett.105.267202} {\bibfield  {journal} {\bibinfo  {journal}
  {Physical Review Letters}\ }\textbf {\bibinfo {volume} {82}},\ \bibinfo
  {pages} {3008} (\bibinfo {year} {2010})},\ \Eprint
  {http://arxiv.org/abs/1008.3507} {1008.3507} \BibitemShut {NoStop}%
\bibitem [{spi()}]{spin_1_2}%
  \BibitemOpen
  \href@noop {} {}\bibinfo {note} {$R_{\mathrm{nuc}}$ is derived for
  spin-$\frac{1}{2}$ particles by replacing $(I_{\pm}^j)^2\rightarrow
  I_{\pm}^j$; this captures the relevant nuclear dynamics.}\BibitemShut {Stop}%
\bibitem [{\citenamefont {Bardou}\ \emph {et~al.}(2003)\citenamefont {Bardou},
  \citenamefont {Bouchaud}, \citenamefont {Aspect},\ and\ \citenamefont
  {Cohen-Tannoudji}}]{levyBook}%
  \BibitemOpen
  \bibfield  {author} {\bibinfo {author} {\bibfnamefont {F.}~\bibnamefont
  {Bardou}}, \bibinfo {author} {\bibfnamefont {J.}~\bibnamefont {Bouchaud}},
  \bibinfo {author} {\bibfnamefont {A.}~\bibnamefont {Aspect}}, \ and\ \bibinfo
  {author} {\bibfnamefont {C.}~\bibnamefont {Cohen-Tannoudji}},\ }\href@noop {}
  {\emph {\bibinfo {title} {{L\'evy Statistics and Laser Cooling}}}}\ (\bibinfo
   {publisher} {Cambridge University Press},\ \bibinfo {year}
  {2003})\BibitemShut {NoStop}%
\bibitem [{t_1()}]{t_1}%
  \BibitemOpen
  \href@noop {} {}\bibinfo {note} {For
  $\Pi_{\mathrm{nuc}}(\omega_{\mathrm{e}}^{\mathrm{eff}})$ to be uniform, the
  distance to the synchronization condition should not exceed a single nuclear
  spin flip $A$, where the waiting time is given by $\tau_1$. Therefore,
  \eqref{eq:waiting} only describes the tail of the distribution, i.e. $\Delta
  t_{\mathrm{nuc}} \gg \tau_1$.}\BibitemShut {Stop}%
\bibitem [{\citenamefont {Mantegna}\ and\ \citenamefont
  {Stanley}(1994)}]{mantegna94}%
  \BibitemOpen
  \bibfield  {author} {\bibinfo {author} {\bibfnamefont {R.}~\bibnamefont
  {Mantegna}}\ and\ \bibinfo {author} {\bibfnamefont {H.}~\bibnamefont
  {Stanley}},\ }\href {\doibase 10.1103/PhysRevLett.73.2946} {\bibfield
  {journal} {\bibinfo  {journal} {Physical Review Letters}\ }\textbf {\bibinfo
  {volume} {73}},\ \bibinfo {pages} {2946} (\bibinfo {year}
  {1994})}\BibitemShut {NoStop}%
\bibitem [{\citenamefont {Erlingsson}\ \emph {et~al.}(2001)\citenamefont
  {Erlingsson}, \citenamefont {Nazarov},\ and\ \citenamefont
  {Fal'ko}}]{erlingsson01}%
  \BibitemOpen
  \bibfield  {author} {\bibinfo {author} {\bibfnamefont {S.~I.}\ \bibnamefont
  {Erlingsson}}, \bibinfo {author} {\bibfnamefont {Y.~V.}\ \bibnamefont
  {Nazarov}}, \ and\ \bibinfo {author} {\bibfnamefont {V.~I.}\ \bibnamefont
  {Fal'ko}},\ }\href
  {https://urldefense.proofpoint.com/v2/url?u=http-3A__journals.aps.org_prb_abstract_10.1103_PhysRevB.64.195306&d=AwIG-g&c=-dg2m7zWuuDZ0MUcV7Sdqw&r=wjB15AxuD9LZg8OWh3q44sP_btMUEdB-Zf79J-wYH_I&m=eGjAXwXFgPKHeLjt85VjZIZzpkxu3N3_2_WSbDCvs6c&s=XDwRI-92jVS0x91L4vHbBsPQLFlj5QcZWwkJ3D2tUk0&e=}
  {\bibfield  {journal} {\bibinfo  {journal} {Physical Review B}\ }\textbf
  {\bibinfo {volume} {64}},\ \bibinfo {pages} {195306} (\bibinfo {year}
  {2001})}\BibitemShut {NoStop}%
\bibitem [{\citenamefont {Dreiser}\ \emph {et~al.}(2008)\citenamefont
  {Dreiser}, \citenamefont {Atat\"{u}re}, \citenamefont {Galland},
  \citenamefont {M\"{u}ller}, \citenamefont {Badolato},\ and\ \citenamefont
  {Imamoglu}}]{dreiser08}%
  \BibitemOpen
  \bibfield  {author} {\bibinfo {author} {\bibfnamefont {J.}~\bibnamefont
  {Dreiser}}, \bibinfo {author} {\bibfnamefont {M.}~\bibnamefont
  {Atat\"{u}re}}, \bibinfo {author} {\bibfnamefont {C.}~\bibnamefont
  {Galland}}, \bibinfo {author} {\bibfnamefont {T.}~\bibnamefont {M\"{u}ller}},
  \bibinfo {author} {\bibfnamefont {A.}~\bibnamefont {Badolato}}, \ and\
  \bibinfo {author} {\bibfnamefont {A.}~\bibnamefont {Imamoglu}},\ }\href
  {\doibase 10.1103/PhysRevB.77.075317} {\bibfield  {journal} {\bibinfo
  {journal} {Physical Review B}\ }\textbf {\bibinfo {volume} {77}},\ \bibinfo
  {pages} {075317} (\bibinfo {year} {2008})}\BibitemShut {NoStop}%
\end{thebibliography}%

\end{document}